\documentstyle[prl,aps]{revtex}
\def\be{\begin{equation}}
\def\ee{\end{equation}}
\def\bea{\begin{eqnarray}}
\def\eea{\end{eqnarray}}
\def\ba{\begin{array}}
\def\ea{\end{array}}

\begin{document}
\title{Octet magnetic moments and the Coleman-Glashow sum rule violation 
in the chiral quark model}
\author{Harleen Dahiya  and Manmohan Gupta}
\address {Department of Physics,
Centre of Advanced Study in Physics,
Panjab University, Chandigarh-160 014, India.} 
 \maketitle

\begin{abstract}
Baryon octet magnetic moments when calculated within the chiral quark
model, incorporating the orbital angular momentum as well as the quark 
sea contribution through the Cheng-Li mechanism, not only show improvement 
over the non relativistic quark model results but also gives a non zero 
value for  the right hand side of Coleman-Glashow sum rule.
When effects due to spin-spin forces between constituent quarks as well as
`mass adjustments' due to confinement are added, it leads to
an excellent fit for the case of $p, \Sigma^+$, $\Xi^o$ and 
violation of Coleman-Glashow sum rule, whereas in almost all
the other cases the results are within 5\% of the data.  
\end{abstract}

The EMC  measurements \cite{EMC} in the deep inelastic scattering
had shown that only a small fraction of the proton's spin is carried by the
valence quarks. This `unexpected' conclusion from the point of view of 
non relativistic quark model (NRQM), usually referred to as 
`proton spin crisis',
becomes all the more intriguing
when it is realized that NRQM is able to give a reasonably
good description of magnetic moments using the assumption
that magnetic moments of quarks are proportional to the spin carried
by them. This issue regarding spin and magnetic moments further becomes 
difficult to
understand when it is realized that the magnetic moments of
baryons receive contribution not only from the magnetic moments carried
by the valence quarks but also from various complicated effects,
such as, orbital excitations \cite{orbitex}, 
sea quark polarization \cite{{eichten},{cheng},{chengsu3},{song}},
effects of the chromodynamic spin-spin forces 
\cite{{Isgur},{mgupta1}}, effect of the confinement
on quark masses \cite{{effm1}}, 
pion cloud contributions \cite{{pioncloudy}}, 
loop corrections \cite{loop},
relativistic and exchange current effects \cite{excurr}, etc.. In the
absence of any consistent way to calculate these effects
simultaneously, even couple of these, it is very difficult to know
their relative contributions. However, the success of NRQM, when viewed
in this context, suggests that the 
various effects mentioned above contribute in a manner where large
part of these is mutually cancelled making the understanding of the
magnetic moments alongwith `spin crisis' all the more difficult. 
The problem regarding magnetic moments gets further 
complicated when one realizes that
Coleman-Glashow sum rule \cite{cg} (CGSR), 
valid in large variety of models
\cite{{cg1},{johan}}, is convincingly violated by the 
data \cite{deltacg}. For example, the CGSR for 
the baryon magnetic moments is given as 
\be
\Delta {\rm CG} = 
\mu_p-\mu_n+\mu_{\Sigma^-}-\mu_{\Sigma^+}+\mu_{\Xi^o}-\mu_{\Xi^-}=0.
\ee
Experimentally
$\Delta{\rm CG}=0.49 \pm 0.05$ \cite{deltacg},
clearly depicting the violation of CGSR by ten standard deviations.
As $\Delta$CG=0, in most of the calculations, obtaining
$\Delta$CG$\neq 0$ alongwith the octet magnetic moments as well as
resolution for `spin crisis' and related issues could perhaps provide vital
clues for the dynamics as well as the
appropriate degrees of freedom required for understanding some of the
non-perturbative aspects of QCD. 

In this context, it is interesting to note that the 
chiral quark model ($\chi$QM) {\cite{{cheng},{manohar}}}
with SU(3) symmetry is not only able to give a fair explanation of
`proton spin crisis' {\cite{EMC}} but is also able to give 
a fair account of $\bar u-\bar d$ asymmetry {\cite{{NMC},{E866},{GSR}}} as well
as the existence of significant strange
quark content $\bar s$ in the nucleon {\cite{st q}}.
Further, $\chi$QM with SU(3) symmetry is also able to provide fairly 
satisfactory explanation for  baryon magnetic moments 
{\cite{{eichten},{cheng},{manohar}}} 
as well as the absence of polarizations of 
the antiquark sea  in the nucleon {\cite{antiquark}} .
The predictions of the $\chi$QM, particularly in regard to  
hyperon decay parameters\cite{decays}, 
can be improved if symmetry breaking effects \cite{{chengsu3},{song}} are
taken into account. However, $\chi$QM with symmetry 
breaking, although gives a fairly good description of magnetic moments, 
is not able to describe $\Delta$CG without resorting to additional
 parameters \cite{{johan},{cgv1}}. 

In a recent interesting  work, Cheng and Li \cite{cheng1} have shown 
that, within  $\chi$QM with SU(3) symmetry breaking, a  long standing
puzzle, ``Why the NRQM
is able to give a fair description of baryon magnetic moments'',
can be resolved if one considers the pions acting as Goldstone Bosons (GBs) 
also have angular momentum and therefore contribute
to the baryon magnetic moments as well. However, this contribution gets
effectively cancelled by the sea quark polarization effect leaving the
description of magnetic moments in terms of the valence quarks in
accordance with NRQM hypothesis. 
One can easily examine that the Cheng and Li proposal does not lead to exact
cancellation of the sea and orbital part for all the baryons,
therefore its implications needs to be examined in detail for the 
octet baryon magnetic moments.
In particular, it would be interesting to explore the possibility of
obtaining non zero $\Delta$CG by invoking Cheng-Li mechanism along with
the mass and coupling breaking effects.

To begin with, we consider the essentials of
$\chi$QM having bearings on the Cheng-Li mechanism. In $\chi$QM, the
basic process is the emission of a GB which further
splits into $q \bar q$ pair, for example,                         

\be
  q_{\pm} \rightarrow GB^{0}
  + q^{'}_{\mp} \rightarrow  (q \bar q^{'})
  +q_{\mp}^{'}.                              \label{basic}
\ee
The effective Lagrangian describing interaction between quarks
and the octet GBs and singlet $\eta^{'}$ is
${\cal L} = g_8 \bar q \phi q,$
where $g_8$ is the coupling constant
and
\[ \phi = \left( \ba{ccc} \frac{\pi^o}{\sqrt 2}
+\beta\frac{\eta}{\sqrt 6}+\zeta\frac{\eta^{'}}{\sqrt 3} & \pi^+
  & \alpha K^+   \\
\pi^- & -\frac{\pi^o}{\sqrt 2} +\beta \frac{\eta}{\sqrt 6}
+\zeta\frac{\eta^{'}}{\sqrt 3}  &  \alpha K^o  \\
 \alpha K^-  &  \alpha \bar{K}^o  &  -\beta \frac{2\eta}{\sqrt 6}
 +\zeta\frac{\eta^{'}}{\sqrt 3} \ea \right). \]

SU(3) symmetry breaking is introduced by considering
different quark masses $m_s > m_{u,d}$ as well as by considering
the masses of GBs to be non-degenerate
 $(M_{K,\eta} > M_{\pi})$ {\cite{{chengsu3},{song},{johan}}}, whereas 
  the axial U(1) breaking is introduced by $M_{\eta^{'}} > M_{K,\eta}$
{\cite{{cheng},{chengsu3},{song},{johan}}}.
The parameter $a(=|g_8|^2$) denotes the transition probability
of chiral fluctuation
of the splittings  $u(d) \rightarrow d(u) + \pi^{+(-)}$, whereas 
$\alpha^2 a$, $\beta^2 a$ and $\zeta^2 a$ 
denote the probabilities of transitions of
$u(d) \rightarrow s  + K^{-(0)}$,
$u(d,s) \rightarrow u(d,s) + \eta$ and
$u(d,s) \rightarrow u(d,s) + \eta^{'}$ respectively.

The magnetic moment corresponding to a given
baryon can be written as
\be
\mu_{total} = \mu_{val} + \mu_{sea} +\mu_{orbit}, \label{totalmag}
\ee
where 
$\mu_{val}=\sum_{q=u,d,s} {\Delta q_{val}\mu_q}$,
$\mu_{sea}=\sum_{q=u,d,s} {\Delta q_{sea}\mu_q}$,    
$\mu_q$ ($q=u,d,s$) is the quark magnetic moment  and $\Delta q$ 
($q=u,d,s$) represents the net spin
polarization  and is defined as
$\Delta q= q_{+}- q_{-}+
\bar q_{+}- \bar q_{-}$ \cite{{chengsu3},{song}}.
The valence contribution $\Delta q_{val}$ can easily be calculated
\cite{{cheng},{chengsu3},{song},{johan}}, for example, for
proton we have $\Delta u_{val} =\left[\frac{4}{3} \right],~~
  \Delta d_{val} =\left[-\frac{1}{3} \right]$ and
  $ \Delta s_{val} = 0$. Similarly one can calculate for other baryons.
 The sea contribution in the $\chi$QM basically comes from the splitting
 of GB into $q \bar q$ pair (Eq (\ref{basic})),
its contribution to baryon magnetic moments can easily be 
calculated within $\chi$QM \cite{{cheng},{song},{johan}}. 
For the case of proton, it is given as
\be
\Delta u_{sea}=- \frac{a}{3} \left (7+4 \alpha^2+
 \frac{4}{3}\beta^2 +\frac{8}{3} \zeta^2 \right),~~
\Delta d_{sea}=-\frac{a}{3} \left(2-\alpha^2
-\frac{1}{3}\beta^2 -\frac{2}{3} \zeta^2 \right),~~
 \Delta s_{sea}=-a \alpha^2.
\ee 
Similarly, one can calculate $\Delta q_{sea}$ and 
consequent contribution to magnetic moments  for other baryons.
Following Cheng and Li  \cite{cheng1}, the $\mu_{orbit}$ for 
$\chi$QM can easily be evaluated and for proton it is given as
\be
\mu_{orbit} =\Delta u_{val} \left[\mu (u_+ \rightarrow) \right]+
\Delta d_{val} \left[\mu (d_+ \rightarrow)
\right],  \label{orbit}
\ee
 where $\mu(u_+ \rightarrow)$ and $\mu(d_+ \rightarrow)$ are the orbital
moments of $u$ and $d$ quarks and are given as
\bea
\mu(u_+ \rightarrow) &=& \frac{a}{2 {M}_{GB}(M_u+{M}_{GB})}
\left[ 3(\alpha^2+1)M_u^2 + (\frac{1}{3}\beta^2+\frac{2}{3}\zeta^2 -
\alpha^2){{M}_{GB}}^2 \right]{\mu}_N, \label{u} \\
\mu(d_+ \rightarrow) &=& \frac{a}{4 {M}_{GB}(M_d+{M}_{GB})} \frac{M_u}{M_d}
\left[ (3- 2 \alpha^2-\frac{1}{3}\beta^2-\frac{2}{3}\zeta^2)
{{M}_{GB}}^2 -6 M_d^2 \right]{\mu}_N,     \label{d} \\
\eea
where $M_q$ and $M_{GB}$ are the masses of quark and GB respectively and
$\mu_N$ is the Bohr magneton. In a similar manner one can calculate the 
contributions for other baryons.

Before discussing  the results,  we would like to discuss the various
inputs pertaining to $\chi$QM which include the mass and
coupling breaking effects. 
The detailed analysis, incorporating the latest E866 data \cite{E866},
to fit quantities such as $\bar u-\bar d$, $\bar u/\bar d$,
$\Delta u$, $\Delta d$, $\Delta s$, $G_A/G_V$, $\Delta_8$, etc. will be
discussed elsewhere, however
here we would like to mention the values of the parameters used for
the calculations of magnetic moments. For example, 
 the  pion fluctuation parameter $a$ 
is taken to be 0.1, in accordance with most of the other
calculations \cite{{eichten},{cheng},{chengsu3},{song},{johan}}, 
whereas the coupling
breaking parameters are found to be $\alpha=0.6$, 
$\beta=0.9$ and $\zeta=-0.3-\beta/2$. 
The orbital angular moment contributions  are  characterized by
the parameters of $\chi$QM as well as the masses of the GBs, however the
contributions are dominated by the pions, therefore the 
effects of other GBs  have been ignored
in the numerical calculations. For evaluating the contribution  
of pions, we have used its on mass shell value in accordance with
several other similar calculations \cite{{mpi1}}.
In the absence of any definite guidelines for the constituent quark
masses, we have used  the most widely accepted values in hadron
spectroscopy 
\cite{{yaouanc},{DGG},{photo},{close}}, 
for example, $M_u=M_d=330$ MeV, whereas
the strange quark mass is taken from the NRQM sum rule
$ \Lambda-N =M_s-M_u$, these quark masses are then used to calculate
$\mu_u$, $\mu_d$ and $\mu_s$.

In Table 1, we have presented the results of our calculations.
From the table, one can easily find out that the $\chi$QM with
symmetry breaking  is not only able to give $\Delta$CG$\neq 0$ but
is also able to give a satisfactory
description of data. When compared with NRQM
results, interestingly one finds that the  results either show improvement
or they have the same overlap with the data as that of NRQM. Specifically
in p, $\Sigma^-$ and $\Xi^0$ one finds that there is a good deal of
improvement as compared to NRQM results. 
In fact, from the table one can easily find out that except
for $\Sigma^-$ and $\Xi^-$, with a marginal correction with
an opposite sign for the latter, in all other cases the sea+orbital
contribution adds to the overall magnetic moments with the
right sign.
These conclusions remain largely valid when one considers variation in
the coupling breaking parameters $\alpha$ and $\beta$. In
particular, for the case of NMC data \cite{NMC}, with different values of
$\alpha$ and $\beta$, we again find 
$\Delta$CG$\neq 0$ as well as  improvement 
over the NRQM results.  The success of $\chi$QM alongwith Cheng-Li
mechanism looks all the more impressive when one realizes that none of the 
magnetic moments have been used as inputs suggesting
that this mechanism seems  to be providing the dominant dynamics of the
constituent quarks and the GBs, the `appropriate' degrees of freedom of
QCD in the non-perturbative regime.

After having seen that $\chi$QM with Cheng-Li mechanism could provide
the dominant dynamics of constituent quarks and the weakly interacting
GBs, it would be natural to consider, as instances of further improvements,
certain effects which can be easily incorporated in $\chi$QM with
Cheng-Li mechanism. To
this end, we have considered  certain phenomenological effects among
constituent quarks 
such as configuration mixing, known to be
compatible with the $\chi$QM  \cite{{prl},{chengspin}},  and the
effects of `confinement', both of which have been shown to improve the
performance of NRQM \cite{{yaouanc},{DGG},{photo},{close}}.  
 For the present purpose, the effect of
configuration mixing on octet baryon wavefunction can be expressed as
{\cite{{mgupta1},{yaouanc}}
\begin{equation}
\left|8,{\frac{1}{2}}^+ \right> = {\rm cos} \phi |56,0^+>_{N=0}
+ {\rm sin} \phi|70,0^+>_{N=2},  \label{mixed}
\end{equation} 
for details of the wavefunction  we refer the reader to Refs.
{\cite{{Isgur},{mgupta1},{yaouanc}}. The angle $\phi$ can be fixed by the 
consideration of neutron charge radius \cite{yaouanc}.
This effectively leads to the change in the valence quark spin
structure, for example, for proton we have $\Delta u_{val} ={{\rm
cos}}^2 \phi \left[\frac{4}{3} \right] 
   + {{\rm sin}}^2 \phi \left[\frac{2}{3}  \right],~~
  \Delta d_{val} ={{\rm cos}}^2 \phi \left[-\frac{1}{3} \right]  +
  {{\rm sin}}^2 \phi \left[\frac{1}{3}  \right]$ and 
  $ \Delta s_{val} = 0$. 
These expressions would replace $\Delta u_{val}$ and $\Delta d_{val}$
in Eq (\ref{orbit}) for  calculating the effects of configuration
mixing on the orbital part. 
Again, for proton,  the sea quark contribution with configuration mixing
can easily be calculated \cite{hd} and is expressed as
\bea
\Delta u_{sea}&=&-{{\rm cos}}^2 \phi \left[\frac{a}{3} (7+4 \alpha^2+
 \frac{4}{3}\beta^2 +\frac{8}{3} \zeta^2)\right]
-{{\rm sin}}^2 \phi \left[\frac{a}{3} (5+2 \alpha^2
+\frac{2}{3}\beta^2 +\frac{4}{3} \zeta^2)\right], \\
\Delta d_{sea}&=&-{{\rm cos}}^2 \phi \left[\frac{a}{3} (2-\alpha^2
-\frac{1}{3}\beta^2 -\frac{2}{3} \zeta^2)\right]
-{{\rm sin}}^2 \phi \left[\frac{a}{3} (4+\alpha^2
+\frac{1}{3}\beta^2 +\frac{2}{3} \zeta^2)\right], \\ 
 \Delta s_{sea}&=&-a \alpha^2.
\eea 
Similarly one can calculate for other
baryons, the details of these calculations will be presented elsewhere.
Before including the effects of configuration mixing on magnetic
moments, because of the changed valence quark spin distribution
functions, one has to carry out a reanalysis to fit 
$\bar u-\bar d$, $\bar u/\bar d$, $\Delta u$, $\Delta d$, $\Delta s$,
$G_A/G_V$, $\Delta_8$ etc. resulting in 
$\alpha=0.4$, $\beta=0.7$ and $\zeta=-0.3-\beta/2$.
Further, apart from taking 
$M_u=M_d=330$ MeV, one has to consider the strange quark
mass implied by the various sum rules derived from the spin-spin
interactions for different baryons {\cite{{mgupta1},{yaouanc}}, for example,
$ \Lambda-N =M_s-M_u$, $(\Sigma^*-\Sigma)/(\Delta-N)=M_u/M_s$ and 
 $(\Xi^*-\Xi)/(\Delta-N)=M_u/M_s$ respectively fix 
$M_s$ for $\Lambda$, $\Sigma$ and $\Xi$ baryons.

It has been shown that the effects of
confinement  can be simulated by `adjusting' the baryon masses 
\cite{effm1}, which leads to the following adjustments for the quark
magnetic moments, for example,
$\mu_d = -\left(1-\frac{\Delta M}{M_B}\right) {\mu}_N,~
\mu_s = -\frac{M_u}{M_s} \left(1-\frac{\Delta M}{M_B}\right) {\mu}_N,
~\mu_u=-2 \mu_d$, where
$M_B$ is the mass of the baryon obtained
additively from the quark masses and $\Delta M$
is the mass difference between the experimental value and $M_B$.

In Table 1, we have included the results obtained after including the
effects of configuration mixing and `mass adjustments'. 
In case of `mass adjustment', one finds a remarkable
improvement for $\Delta$CG due to the difference in the valence 
contributions getting affected with the right magnitude, however 
the individual magnetic moments
get disturbed. Interestingly the situation changes
completely when configuration mixing is also included. From  the
table, it is evident that we are able to get an excellent fit for
almost all the baryons, it is almost perfect for $p, \Sigma^+$ and $\Xi^o$,
in the case of $n, \Sigma^-$ and $\Lambda$ 
the value is reproduced within 5\% of
experimental data. Only in the case of $\Xi^-$ the deviation is
somewhat more than 5\%. 
Besides this we have also been able to get an excellent fit to 
$\Delta$CG. 
The fit becomes all the more impressive when it is
realized that none of the magnetic moments are used as inputs. 
It may be of interest to mention that the fit in the case of $\Xi^-$
can perhaps be improved in the present case if corrections due to
 pion loops are included \cite{pioncloudy}. In fact, a cursory look at
Ref. \cite{pioncloudy} suggests that pion loop corrections would compensate
$\Xi^-$ much more compared to other baryons hence providing an almost
perfect fit.

To summarize our conclusions, we have carried out a detailed
calculation of the octet magnetic moments in the $\chi$QM by including
orbital and sea 
contributions through Cheng-Li mechanism \cite{cheng1} with
mass and coupling breaking effects. Apart from reproducing
$\Delta$CG$\neq 0$, we are able to show that Cheng-Li mechanism is able
to improve the results of NRQM without any additional inputs
indicating that this mechanism provides the dominant dynamics of the
constituent quarks and the Goldstone Bosons. This is further borne out by the
fact that when effects such as configuration mixing due to spin-spin
interactions and `mass adjustments' due to confinement are added, we
are able to get an excellent
fit for the octet baryon magnetic moments as well as for $\Delta$CG, it
is almost perfect for $p, \Sigma^+$ and $\Xi^o$ whereas
in the case of $n, \Sigma^-$ and $\Lambda$
the value is reproduced within 5\% of
experimental data. 

The authors would like to thank S.D. Sharma for fruitful 
discussions. H.D. would like to thank CSIR, Govt. of India, for
financial support and the Chairman, Department of Physics, 
for providing facilities to work in the department.

\pagebreak

\begin{table}[h]
{\footnotesize
\begin{center}
\begin{tabular}{ccccccccccccccc}      
 &  & & \multicolumn{4}{c} { } & \multicolumn{4}{c} {} & 
\multicolumn{4}{c} {$\chi$QM with mass }\\  

&  & & \multicolumn{4}{c} {$\chi$QM} & \multicolumn{4}{c}
{$\chi$QM with} &  \multicolumn{4}{c} {adjustments and}\\ 
&  & & \multicolumn{4}{c} {} & \multicolumn{4}{c}
{mass adjustments} &  \multicolumn{4}{c} {configuration
mixing}\\ 
\cline{4-7}  \cline{8-11} \cline {12-15}

Octet & Data & NRQM &  Valence  & Sea & Orbital &  Total &  
Valence  & Sea & Orbital &  Total &  Valence  & Sea & Orbital &  Total \\
 baryons & ~\cite{deltacg} &  &   & &&&&&&&&&& \\  \hline

p          & 2.79  & 2.72 & 3.00 & -0.70 & 0.54 & 2.84 &  
3.17   & -0.59&    0.45 & 3.03 &  
2.94    & -0.55& 0.41 & 2.80   \\
n          & -1.91 & -1.81 & -2.00 & 0.34 & -0.41 & -2.07 & 
-2.11    & 0.24  & -0.37 & -2.24 & 
 -1.86  &  0.20 & -0.33  & -1.99   \\

$\Sigma^-$ & -1.16 &  -1.01 & -1.12 & 0.13 & -0.29 & -1.28 & 
-1.08 & 0.08& -0.26 &   -1.26 &
   -1.05    & 0.07 &  -0.22 & -1.20  \\
$\Sigma^+$ & 2.45  & 2.61 & 2.88 & -0.69 & 0.45 & 2.64 &  
2.80  &  -0.55 &  0.37 & 2.62 & 
 2.59 & -0.50 & 0.34 &  2.43  \\

$\Xi^o$    & -1.25 & -1.41 & -1.53 & 0.37 & -0.23 & -1.39 & 
-1.53 & 0.22 &  -0.16 &  -1.47 & 
 -1.32 & 0.21 &-0.13 & -1.24  \\
$\Xi^-$    & -0.65 & -0.50 & -0.53 & 0.09 & -0.06 & -0.50 & 
-0.59 & 0.06  & -0.01 & -0.54 &  
 -0.61 & 0.06&  -0.01 &   -0.56  \\

$\Lambda$ & -0.61  & -0.59 & -0.65 & 0.10 & -0.08 & -0.63 & 
-0.69  & 0.05  & -0.04  &-0.68  &
-0.59 & 0.04 &-0.04  & -0.59   \\
 \hline
$\Delta$CG  & 0.49 $\pm$ 0.05 & 0 &  & & & 0.10 & 
& &&  0.46 &  & &&     0.48 \\ 

\end{tabular}
\end{center}}
\caption{ Octet baryon magnetic moments in units of $\mu_N$. The
details of inputs are discussed in the text.}  
\end{table}

\end{document}